\documentstyle[aas2pp4]{article}

\def\deg{\ifmmode^\circ\else$^\circ$\fi}

\def\Msol	{M$_{\odot}$}
\def\Lsol	{L$_{\odot}$}

\def\gtsim	{\ {\raise-0.5ex\hbox{$\buildrel>\over\sim$}}\ }
\def\ltsim	{\ {\raise-0.5ex\hbox{$\buildrel<\over\sim$}}\ }
\def\sqdeg{$\Box$\deg}
\received{18 July 1997}
\slugcomment{Submitted to {\em The Astrophysical Journal Letters }.}
\lefthead{Tinney et al.}
\righthead{DENIS-P\,J1228.2-1547 -- A New Benchmark Brown Dwarf}

\begin{document}
\title{DENIS-P\,J1228.2-1547 -- A New Benchmark Brown Dwarf }

\author{C.G. Tinney}
\affil{Anglo-Australian Observatory, PO Box 296, Epping. 2121. \\
Australia. email: cgt@aaoepp.aao.gov.au}
\author{X. Delfosse and T. Forveille}
\affil{Observatoire de Grenoble,
              414 rue de la Piscine,\\
              Domaine Universitaire de S$^{\mathrm t}$ Martin d'H\`eres,
              F-38041 Grenoble,
              France}

\begin{abstract}
We present optical spectroscopy of three brown dwarf candidates identified in
the first 1\% of the DENIS sky survey. Low resolution 
spectra from 6430--9150\,\AA\ show these objects to have similar spectra to the 
brown dwarf candidate GD\,165B. High resolution spectroscopy shows that 
one of the objects -- DENIS-P\,J1228.2-1547 -- has a strong 2.3$\pm$0.05\,\AA\  
equivalent width absorption line of Li I 6708\,\AA, and is therefore a 
brown dwarf with mass below 0.065\,\Msol, and age $\ltsim 1.5$\,Gyr.
DENIS-P\,J1228.2-1547 can now be considered a proto-type 
for brown dwarfs closer to the hydrogen burning limit than Gl\,229B.
\end{abstract}


\section{Introduction}

Claims of brown dwarf discovery have historically been problematic. Numerous 
objects have been claimed as brown dwarfs or brown dwarf candidates over the
last 15 years, but few have been either generally accepted or subsequently
confirmed. This situation changed -- and in spectacular fashion -- with the 
discovery of the brown dwarf Gl\,229B (\cite{nak+95} 1995). Here at last was a brown
dwarf proto-type so cool, and of such low luminosity, that it quickly became
universally accepted. Unfortunately, what Gl\,229B didn't resolve was the status
of more massive brown dwarf candidates.

For example, the nature of GD\,165B  
-- the next best old brown dwarf candidate -- remains unclear 
(\cite{bz88} 1988; \cite{kir97a} 1997a). Although its
optical spectrum is qualitatively different from that of low-mass stars
(\cite{khl93} 1993; \cite{kbs97} 1997), its infrared spectrum is not 
(\cite{jljm94} 1994). And unfortunately, the closeness of its
white dwarf companion makes the detection of Li almost impossible. 
\cite{tjh97} (1997) have found a likely field brown dwarf (296A),
but it is of {\em very} early spectral type (M5.5), indicating an age of less than
200\,Myr. There is a  
need for a late proto-type object near the star-to-brown-dwarf transition, but which 
is {\em clearly} a brown dwarf, against which objects like GD\,165B can be
compared. We present here observations of just such an
object, DENIS-P\,J1228.2-1547, discovered in the first $\sim$1\% of data examined
from the Deep Near Infrared Southern sky survey (DENIS).

\section{Observations}

The DENIS survey will cover the entire southern sky in three infrared pass bands to 3-$\sigma$ limits of I=18.5, J=16.5, K$^\prime$=13.5 (\cite{e97} 1997; \cite{cop+97} 1997). Such a survey is ideally suited to finding field brown dwarfs. 
The DENIS mini-survey project (\cite{dtf+97} 1997; \cite{dtf97} 1997) has begun 
this search by observing 
brown dwarf candidates from the first $\sim$1\% (230\,\sqdeg) of the 
initial DENIS data. Infrared 
spectroscopy obtained on the 3.9m Anglo-Australian Telescope (AAT) confirmed that at 
least three of the mini-survey objects were as cool, or cooler than, GD\,165B (\cite{dtf+97} 1997).

Optical spectroscopy was obtained with the AAT on 
1997 June 7-9 (UT), using the RGO Spectrograph with TEK 1K CCD\#2. Observations
 were made using both a 270R grating in blaze-to-camera mode, providing a
 resolution of 7\,\AA\ and a wavelength coverage of 6425--9800\,\AA, and a 
1200R grating in blaze-to-collimator mode giving a resolution of 1\,\AA\ and 
a wavelength coverage of 6495--7030\,\AA. The latter set-up was specifically 
chosen to permit observation of both the H$\alpha$ and Li\,I~6708\,\AA\ lines. 
All three DENIS objects of interest, as well as several 
very low-mass stars were observed. The observations are summarised in Table 1. 
Finding charts for the three DENIS objects can be found in 
\cite{dtf+97} (1997). These data were processed using standard techniques within 
the FIGARO data reduction package (\cite{sho93} 1993). 

%\placetable{tbl-1}

\begin{deluxetable}{lccc}
\footnotesize
\tablecaption{Observation Log. \label{tbl-1}}
\tablewidth{0pt}
\tablehead{
\colhead{Object} & \colhead{Position$^a$}   & \colhead{270R Exp.}   & \colhead{1200R Exp.} \\
\colhead{ }      & \colhead{(J2000.0)}   & \colhead{(s)}   & \colhead{(s)} 
} 
\startdata
DENIS-P\,J1228.2-1547     & 12:28:13.8  $-$15:47:11& 1800    & 18000 \\ % 19
DENIS-P\,J1058.7-1548     & 10:58:46.5  $-$15:48:00& 3600    &  7200 \\   % 22
DENIS-P\,J0205.4-1159     & 02:05:29.0  $-$11:59:25& 5400    & --     \\       % 37
BRI\,0021-0214      & 00:24:24.6  $-$01:58:22& 1800    &  1800 \\
VB\,10/LHS~474       & 19:16:57.9  $+$05:09:10&  600    & --     \\
 \enddata
\tablenotetext{a}{Positions for the DENIS objects are from \cite{dtf+97} 1997, those for the remainder are from \cite{trgm95} 1995. The DENIS-P prefix indicates that these are
provisional DENIS objects, which have not been produced by the final DENIS  
catalogue pipeline.}
 
\end{deluxetable}

\section{Discussion}

The resulting spectra are shown in Figure 1 (the low resolution spectra) 
and Figure 2 (the high resolution spectra). Perhaps the most instantly 
noticeable feature of Figure 1 is the remarkable similarity between the
spectra of the three DENIS objects and the spectra of GD\,165B -- for so 
long seen as a singleton object which fits into no classification 
scheme (Kirkpatrick et al. 1997; Kirkpatrick et al.  1993).

%\placefigure{fig1}

\begin{figure*}
\epsscale{1.85}
\plotone{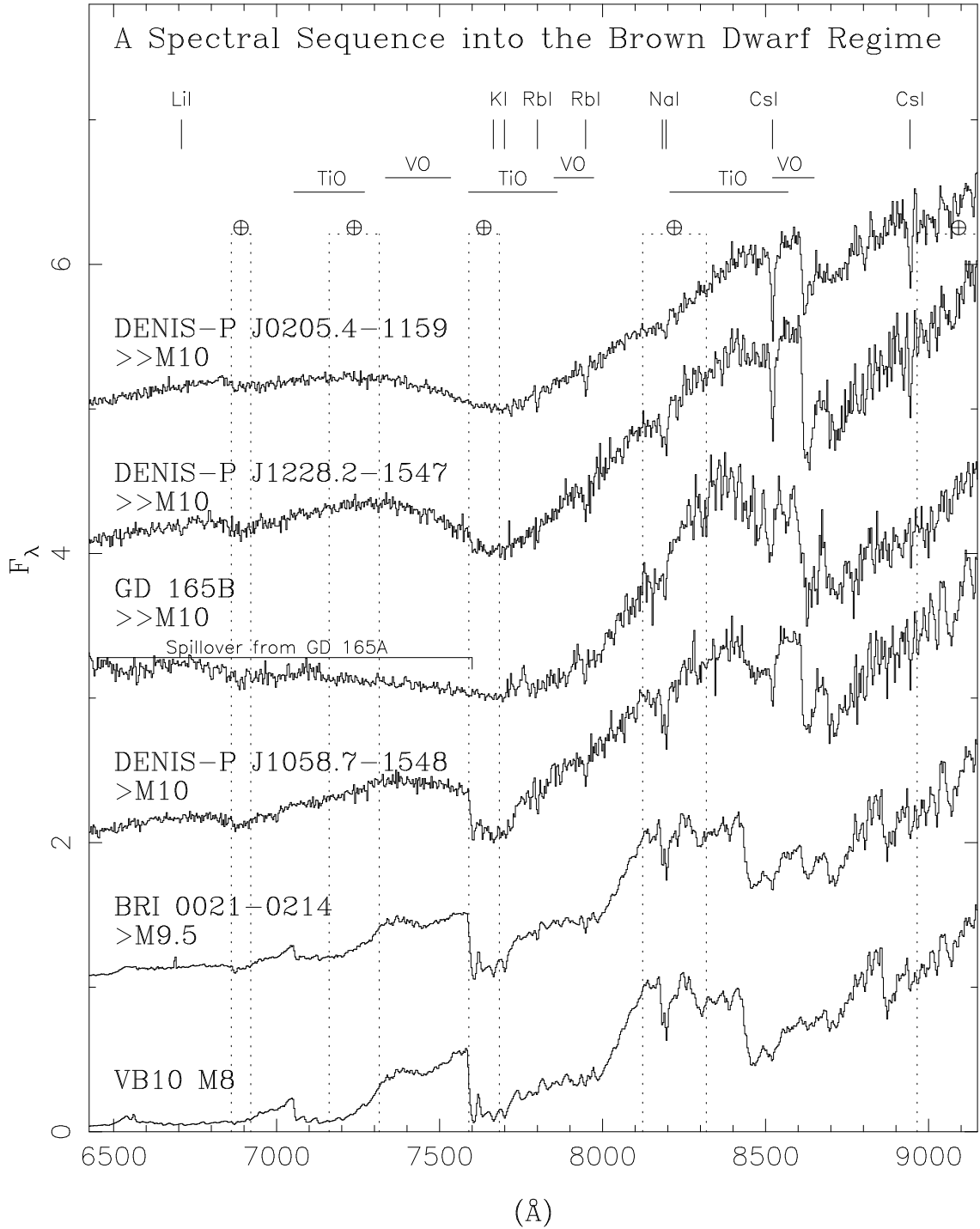}

\caption{Low resolution (7\,\AA) AAT spectra of the DENIS brown dwarf candidates (and two comparison spectra) arranged in approximate order of spectral lateness, together with
the GD\,165B spectrum of Kirkpatrick et al. (1993), which is subject to
contamination below 7600\,\AA. Each spectrum has been normalised to unity at 8800\,\AA, 
and offset in unit steps for clarity. The spectral types shown are those due 
to Kirkpatrick et al. (1997), or our estimates on their system.  Stellar atomic and molecular absorption features are marked, as are terrestrial absorption features, which have not been corrected. \label{fig1}}
\end{figure*}

\subsection{Low-resolution spectra}

The spectra in Figure 1 have been organised in their apparent order of 
spectral type -- latest towards the top. Because of the strength of terrestrial
H$_2$O, the spectra are not plotted beyond 9150\,\AA.
The spectral types shown for VB\,10,
BRI\,0021-0214 and GD\,165B are due to Kirkpatrick et al. (1997) -- 
the other types are our estimates.
 Prominent in the spectra are the lines of Cs I at 8521\,\AA\ and 8943\,\AA.
 These lines were pointed out in the spectra of very low-mass stars by 
\cite{tin97} (1997), at equivalent widths (EW) of 0.5-1\,\AA. In the latest of the DENIS 
objects (DENIS-P\,J1228.2-1547 and DENIS-P\,J0205.4-1159) they are present at a whopping 
EW=6.5\,\AA\footnote{Whenever equivalent widths are referred to 
they are always the ``psuedo-equivalent width'' defined by the apparent 
continua available near the lines, at the resolution of the observations. 
In practice, no part of the spectrum of these objects is free of molecular 
absorption, so no absolute continuum is ever available.}. Also strong are 
lines of Rb~I at 7800\,\AA\ and 7943\,\AA\ (\cite{bm95} 1995; \cite{tin97} 1997). Interestingly the lines of Na~I and K~I, so prominent in late M-dwarfs, 
seem to become progressively weaker beyond M10.

However, the strongest feature of Figure 1 is the almost counter-intuitive 
weakening of the TiO and VO molecular bands with decreasing temperature. 
This is primarily due to the formation of dust, which plays an important 
role in these very late atmospheres. This is both because
of its influence on opacities (eg. models require opacity due to dust to match 
the spectral energy distribution of GD\,165B -- \cite{toan96} 1996) {\em and} composition 
(eg. the formation of perovskite, CaTiO$_3$, will deplete atmospheres at 
$\sim$ 1600K of their TiO, causing these bands to vanish --
\cite{all97} 1997; \cite{kir97a} 1997a). Notice in particular the bands near 8500\,\AA\ which
are weakly present in GD\,165B, but have vanished at the temperatures
of DENIS-P\,J1228.2-1547 and DENIS-P\,J0205.4-1159.

The spectra confirm the result of \cite{dtf+97} (1997) -- these three 
objects have effective temperatures similar to, or cooler than, GD\,165B. 
Like GD\,165B, the three DENIS objects all show weak, or non-existent, TiO and VO 
bands, as well as strong lines of Cs I (\cite{kir97b} 1997b). 
In fact, apart from being very red, their spectra look almost nothing like 
those of late M dwarfs. This supports the suggestion of \cite{kir97a} (1997a) that
an entirely new spectral type is required for this class of objects --  
based on features like the absence of TiO and VO, and the strength of Cs~I.

%\placefigure{fig2}

\begin{figure}
\epsscale{1.0}
\plotone{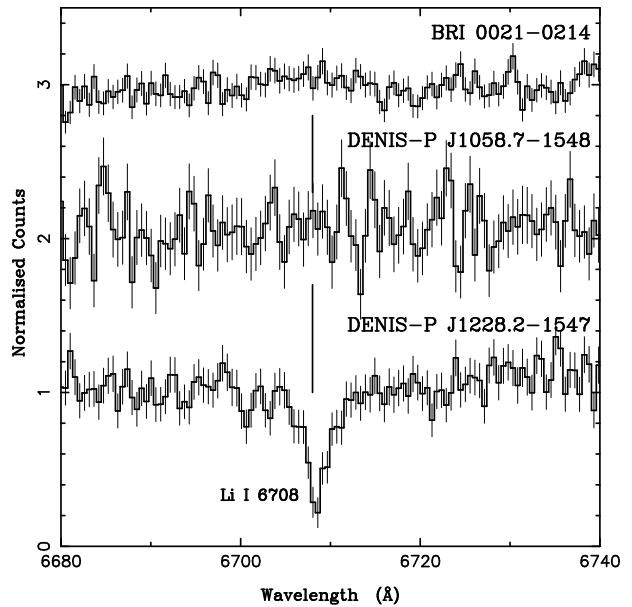}
\caption{High resolution (1\,\AA) AAT spectra in the region of the 
Li\,I~6708\,\AA\ line. Each spectrum has been normalised to unity at 6660-6680\,\AA, and  offset in unit steps for clarity. Error bars 
show the propogated photon-counting uncertainties. \label{fig2}}
\end{figure}

\subsection{Lithium Spectra}

Figure 2 shows the high resolution spectra obtained for DENIS-P\,J1228.2-1547 
and DENIS-P\,J1058.7-1548, as well as a comparison spectrum of BRI\,0021-0214, 
which is known to have depleted its lithium (\cite{bm95} 1995). 
DENIS-P\,J1228.2-1547 shows a clear Li
absorption feature with EW=2.30$\pm$0.05\,\AA. (The quoted uncertainty is based
on photon-counting errors, and does not include the systematic uncertainty 
in the ``pseudo'' equivalent width.) Despite the 
fact that DENIS-P\,J1058.7-1548 was observed for only 2 hours (as compared 
to 5 hours  for  DENIS-P\,J1228.2-1547) it clearly {\em does not} show a Li absorption of a  similar strength. We place a 3$-\sigma$ upper limit 
on this object of EW$<$0.5\,\AA.

Interpretation of these equivalent widths as Li abundances is made problematic
by the lack of published model atmospheres at the required spectral resolution 
and temperatures. However, based on the models and curves of growth of \cite{prmgl95} (1995) which extend to 2000K, 
we crudely estimate [Li/H] = $\sim$ 1--3. 
\cite{dtf+97} (1997) estimate absolute magnitudes for
DENIS-P\,J1228.2-1547 and DENIS-P\,J1058.7-1548 of M$_{K}$ = 12.1$\pm$0.4 and 11.4$\pm$0.4 (respectively). These seem to be confirmed by the Fig. 1 spectra, which show
DENIS-P\,J1058.7-1548 being of earlier type than GD\,165B (M$_{K}$ = 11.7$\pm$0.2; \cite{dahn97} 1997; \cite{tmr93} 1993), and DENIS-P\,J1228.2-1547 being of later type.
Bolometric corrections have been estimated for only one object
at these temperatures (GD\,165B, \cite{tmr93} 1993), however the unceratinties
in BC$_K$ are, in any case, considerably smaller than those in M$_{K}$.
We derive M$_{bol}$ = 15.4$\pm$0.5 and 14.7$\pm$0.5, or log(L/\Lsol)=$-$4.3 and $-$4.0. At log(L/\Lsol) = $-$4.3, the presence of {\em any lithium at all} in DENIS-P\,J1228.2-1547 
constrains the mass to be less than 0.065\,\Msol, and the
age to be $\ltsim 1$\,Gyr (\cite{nrc93} 1993). It is clearly a {\em bona fide}
brown dwarf -- it cannot be a more massive object which has not yet depleted
its lithium, since it is more than 10 times fainter than models at the required age ($\sim$ 100\,Myr). Conversely, DENIS-P\,J1058.7-1548 must be more massive than 0.065\,\Msol.

Recently, an improved understanding of convection in cool atmospheres has found that
for T$_{eff} \leq 2200$K, the central convection zone does {\em not} reach into
the photosphere (\cite{all97} 1997). This lead Allard to suggest that the ``Li test'' 
is inappropriate for M10 dwarfs and later. While this is true, objects more
massive than 0.05\,\Msol\ will spend at least $\approx$3$\times$10$^8$\,yr at
T$_{eff}$$\ge$2300K (\cite{cbp96} 1996; \cite{bhsl93} 1993), during which time their Li will be depleted
if central temperatures are high enough (ie if their mass is $\geq$0.065\,\Msol), and
will not be depleted if their mass is $\leq$0.065\,\Msol. For objects less massive than
0.05\,\Msol\ the presence of Li will reflect both the initial Li abundance, and the
timescale at which the convective interior drops below the photosphere. In either case,
the presence of Li in DENIS-P\,J1228.2-1547 must imply a brown dwarf nature, since Li would
have been totally depleted in an object with its luminosity were it more massive than 0.065\,\Msol.

\subsection{H$\alpha$ emission}

Interestingly, only one of the DENIS objects reported here shows evidence for H$\alpha$ emission -- and that is very weak.  Our high resolution spectra show H$\alpha$ 
emission with EW=1.3$\pm$0.4\,\AA\ in DENIS-P\,J1058.7-1548.  We place 3-$\sigma$ 
upper limits on emission in DENIS-P\,J1228.2-1547 and DENIS-P\,J0205.4-1159 of 1.0\,\AA\ and 
3.5\,\AA, respectively. The level of emission in these, presumably old, dwarfs is
significantly less than that seen in the Pleiades brown dwarfs (age $\sim$ 100\,Myr, EW$\approx$5\,\AA\, \cite{rmb+96} 1996). In particular, this implies that DENIS-P\,J1228.2-1547
probably has an age $\gtsim 100$\,Myr. Given its luminosity, this imposes a lower mass
limit of about 0.02\,\Msol (\cite{bhsl93} 1993).

We also detect H$\alpha$
emission at a level of 1.30$\pm$0.05\,\AA\ in BRI\,0021-0214, in which H$\alpha$ had
not previously been detected (\cite{bmg96} 1996). This clearly indicates that
chromospheric activity {\em is} present in this star, though at a low level.

\section{Conclusions}

Finally a brown dwarf has been discovered, just below the
brown dwarf limit, conclusively showing us what a brown dwarf of 
$\ltsim$0.065\,\Msol\ looks like. The similarity of the optical and infrared
spectra of DENIS-P\,J1228.2-1547 to that of GD\,165B, and the other two DENIS objects
presented here, suggests that they too probably lie below the brown dwarf
limit. In fact, the available data point to DENIS-P\,J0205.4-1159 being {\em cooler}
than DENIS-P\,J1228.2-1547 -- Li observations of this object will be made a high
priority. Lastly, it is worth emphasising that these three dwarfs were
discovered in just the first $\sim$1\% of the DENIS data -- many more
exciting results can be expected in the future as this survey continues.

\acknowledgments

We are grateful to Davy Kirkpatrick
for discussions on his unpublished GD\,165B data and for a thorough
referee's report, and to Ben Oppenheimer for helpful
discussions about his unpublished Gl\,229B data. We are 
also extremely grateful to the AAO Director, Brian Boyle, for providing 
the discretionary time in which these observations were carried out.
CGT would personally like to thank Nicolas Epchtein and the members of the DENIS
team for providing access to DENIS data prior to its general release. The 
DENIS project is partly funded by the European Commission through the
{\em SCIENCE} and {\em HCM} grants. It is also supported, in France by INSU,
the Education Minsitry and CNRS; in Germany by the Land of Baden-W\"urtenburg;
in Spain by DGICYT; in Italy by CNR; in Austria by the FFWF and BWF; in
Brazil by FAPESP; and in Hungary by an OTKA grant. It is also partly supported
by an ESO C\&EE grant.

%\clearpage

%\plotone{tdf_li_fig1.ps}
%
%\plotone{tdf_li_fig2.ps}

\end{document}